\documentclass[twocolumn,amsmath,amssymb,aps,pra,floatfix,showpacs]{revtex4-1}
\usepackage{graphicx}
\usepackage[capitalise]{cleveref}
\begin{document}
\title{Ultrafast dynamic evolution of multilevel systems in medium-strength laser fields}
\author{Zhenhao Wang}
\author{Yu He}
\author{Quanjun Wang}
\author{Jingjie Ding}
\author{Shaohua Sun}
\author{Zuoye Liu}
\email{Corresponding author: zyl@lzu.edu.cn}
\author{Bitao Hu}
\affiliation{School of Nuclear Science and Technology, Lanzhou University, Lanzhou 730000, China}
\date{\today}

\begin{abstract}
The ultrafast dynamic evolution of an atomic system under medium-strength laser fields is studied by performing transient absorption measurement. An analytical model developed from perturbation theory with a modified transition dipole moment is presented to explain the spectral features of the multilevel system. By fitting the measured absorption spectra to the model, the system's dynamic evolution is quantified by different amplitude and phase modulation factors in the pump--probe and probe--pump scenarios. This study provides a way to understand laser--matter interaction in the transition area between the strong-field and weak-field regimes.
\end{abstract}

\maketitle

\section{Introduction}

Laser--matter interaction is studied by charged particle detection~\cite{Momentumspectroscopy,Charged_particle_detection_electron2,Charged_particle_detection_electron1,Charged_particle_detection_electron_coincidence,Charged_particle_detection_electron3} and light measurement~\cite{Chem.Phys.Lett.133.373, doi:10.1002/9780470259498.ch6,pump_probe1,pump_probe2,Nature.434.625,Proc.Natl.Acad.Sci.USA.107.12766}. Detection of the generated charged particles requires the release of photoelectrons and thus is suitable only for the study of ionizing events. In contrast, light measurement is not limited to phenomena that liberate electrons; it is the method of choice for investigating bound--bound transitions with a higher energy resolution than ion or electron detection~\cite{chargedparticle}. Among these spectroscopic methods, pump--probe spectroscopy~\cite{pump_probe1,pump_probe2} revolutionizes our understanding of electron dynamics by capturing ultrafast processes in atoms, molecules, and solids in real time. The radiation of a sample's response is self-heterodyned in the pump--probe geometry, and the temporal evolution of the target system manifests itself as the relative intensities of light absorbed at different frequencies. Traditionally, weak laser fields are applied in pump--probe spectroscopy, and the signal is analyzed by third-order perturbation theory~\cite{PeterHammSpec}, yielding an intuitive description of light--matter interaction. However, the perturbative description is no longer valid when the laser fields are sufficiently intense to remove a significant fraction of the population from the ground state. Furthermore, the AC Stark effect induced by these intense laser fields will profoundly modify the energy-level structure of the sample, and the population of the upper state in a two-level system no longer follows the Fermi golden rule~\cite{Allen1975Optical} but is governed by Rabi oscillation~\cite{BoydNonlinearoptics,Rabiosc}.

Recently, it was demonstrated that the phase of a sample's polarization induced by a resonant laser field in an atomic system can be modified by a strong nonresonant laser field owing to the AC Stark effect~\cite{fanoprofile,strongpump1,strongpump2,strongpump3}. This model is known as the dipole phase control theory and has been extended to multilevel and complex molecular systems by using strong pump and weak probe pulses with the same frequency~\cite{Proc.Natl.Acad.Sci.USA.112.15613,spspLiu1,spspLiu2}. Actually, both the phase and amplitude of a sample's response in the time domain will be significantly modified by the AC stark effect and Rabi oscillation, respectively, regardless of whether the probe pulse precedes or follows the pump pulse. This feature can be used to obtain a fully tailored response from the target system. For laser field intensities too high for perturbation theory to be used, numerical calculations were done to understand the novel behavior of the sample due to the high nonlinearity of light--matter interaction~\cite{spspGelin1,spspGelin2,spspGelin3,spspGelin4}. However, this method gives no insight into the general properties of the interaction; thus, analytical models are needed to understand the corresponding dynamics.

In this paper, a novel approach is developed to characterize the behavior of the atomic system in medium-strength laser fields, which is known as the transition area between strong and weak fields. The effect of Rabi oscillation, weak transitions, and higher-order interactions cannot be ignored in this regime. They are considered as a modification of third-order perturbation theory and inherit the valuable characteristics of perturbation theory in the medium-strength laser field regime. This approach is described using the double-sided Feynman diagram (DSFD)~\cite{PeterHammSpec} and is then applied to explain the time evolution of a three-level V-type system interacting with medium-strength laser fields. Instead of the pump-pulse intensity being changed, the probe-pulse intensity is tuned to examine the strong field effect on the dynamic evolution of the target multilevel system.

\section{Analytical model of V-type three-level system in medium-strength laser fields}

In the pump--probe geometry, the probe pulse has a time delay $\tau$ with respect to the pump pulse. Positive time delays correspond to the typical pump--probe scenario, in which the probe pulse arrives after the pump pulse to interact with the sample, whereas negative time delays correspond to the probe--pump scenario. It is assumed that the semi-impulse limit is satisfied, i.e., the interaction is faster than the dephasing time of the sample but slower than the oscillation period of the laser field, so the envelope of the laser pulse can be approximated as a $\delta$ function while the oscillating field phase factor is retained. A phenomenon called a coherence spike or coherent artifact occurs~\cite{coherence_spike1,coherence_spike2,coherence_spike3} during the overlap of the pump and probe pulses but will not be considered in this work. In the weak-field regime, the third-order polarization based on perturbative expansion of the Liouville–-von Neumann equation in the interaction picture reads as~\cite{PeterHammSpec}
\begin{equation}
\begin{split}
P^{(3)}(t)\propto&\int_0^\infty\int_0^\infty\int_0^\infty dt_1dt_2dt_3 R(t_3,t_2,t_1)E_3(t-t_3)\\
&E_2(t-t3-t2)E_1(t-t_3-t_2-t_1)\\
\propto& R(t_3,t_2,t_1),
\end{split}
\end{equation}
where $R(t_3,t_2,t_1) \propto$
\begin{equation}
\begin{split}
-(-\frac{i}{\hbar})^3\langle\hat{\mu}(t_3+t_2+t_1)[\hat{\mu}(t_2+t_1),[\hat{\mu}(t_1)[\hat{\mu}(0),\rho(-\infty)]]]\rangle
\end{split}
\label{equ:reponse function}
\end{equation}
is the total response function, including all possible pathways of the dynamic evolution of the sample. Here, angle brackets indicate the trace. The first, second, and third interactions, which are induced by laser fields $E_1$, $E_2$, and $E_3$, occur at times $t=0$, $t=t_1$, and $t=t_1+t_2$, respectively. In the medium-strength-field regime, a modification of the transition dipole moment (TDM) $\hat{\mu}(t)$ in \eqref{equ:reponse function} is introduced to account for the Rabi oscillation and contributions from  high-order interactions, higher excited states, and weak transitions that are not included in the model system. The modification is considered as transient under the semi-impulse limit.

\begin{figure}[ht]
\centering
\includegraphics[width=3in]{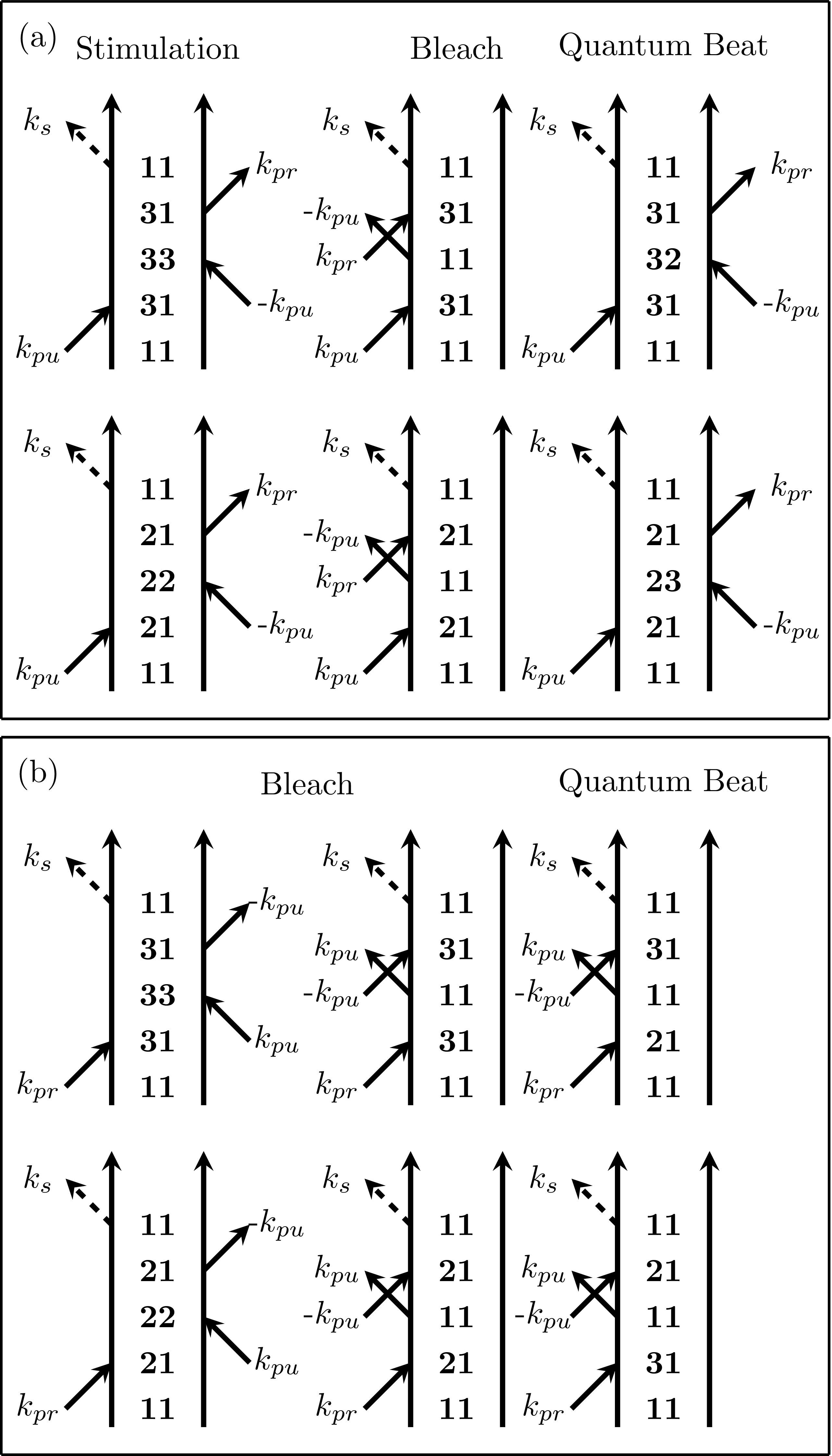}
\caption{Survival DSFDs of V-type three-level system in pump--probe spectroscopy for (a) the pump--probe scenario and (b) the probe--pump scenario. The pump--probe scenario has stimulation, bleach, and quantum beat pathways, and the probe--pump scenario has only the bleach and quantum beat pathways.}
\label{fig:feynman}
\end{figure}

To describe the modification clearly, an analytical model of the time evolution of a V-type three-level system in medium-strength laser fields is developed using the DSFD. The three-level system is composed of the ground state $|1\rangle$ and the excited states $|2\rangle$ and $|3\rangle$. Electric dipole transitions $|1\rangle \rightarrow |2\rangle$ with frequency $\omega_{21}$ and $|1\rangle \rightarrow |3\rangle$ with frequency $\omega_{31}$ are permissible, whereas $|2\rangle \rightarrow |3\rangle$ is forbidden. Further, $\mu_{21}(t)$ ($\mu_{31}(t)$) is the TDM of transition $|1\rangle \rightarrow |2\rangle$ ($|1\rangle \rightarrow |3\rangle$). The DSFDs meeting the rotating wave approximation and phase matching for the pump--probe and probe--pump scenarios are shown in Figs.\,\ref{fig:feynman}(a) and \ref{fig:feynman}(b), respectively. The bleach and quantum beat~\cite{quantumbeat} pathways exist in both the pump--probe and probe--pump scenarios, whereas the stimulation pathway cannot be found in the probe--pump scenario, because the radiation of this pathway has the same direction as the pump pulse. Among these possible pathways, the stimulation pathway reduces the absorption of the probe pulse, whereas the bleach pathway has the opposite effect, and the quantum beat pathway is caused by coherence between transitions $|1\rangle \rightarrow |2\rangle$ and $|1\rangle \rightarrow |3\rangle$. When medium-strength laser fields are applied, the modification of the system's response can be intuitively formalized by the following method. For the interactions depicted by arrows in the DSFDs, except for the last arrow originating from the trace $P^{(3)}(t)=\langle\hat{\mu}\rho^{(3)}(t)\rangle$, each arrow introduces a complex modification factor $\sqrt{a}e^{i\phi}\,(a>0)$ into the TDM of the corresponding transition, i.e., $\mu_{mn}(t)$ for transition $|n\rangle\rightarrow |m\rangle$ is modified as $\mu_{mn}(t)\sqrt{a}e^{i\phi}$ during the interaction. An arrow pointing toward (away from) the left side of the DSFD introduces a positive (negative) phase modification; in contrast, an arrow pointing toward (away from) the right side introduces a negative (positive) phase modification. 

In the pump--probe scenario, two interactions are induced by the pump pulse for the stimulation pathway in Fig.\,\ref{fig:feynman}(a). The interaction at time $t=0$ points toward the left side of the DSFD, whereas the interaction at $t=t_1$ points toward the right side. The interaction induced by the probe pulse points away from the right side of the DSFD. The last interaction, depicted by a dotted arrow, will not be modified. Therefore, the TDM $\mu_{21}(t)$ becomes $\mu_{21}(0)\sqrt{a_{21}^{\rm pu}}e^{i\phi_{21}^{\rm pu}}$ at $t=0$, $\mu_{21}(t_1)\sqrt{a_{21}^{\rm pu}}e^{-i\phi_{21}^{\rm pu}}$ at $t=t_1$, and $\mu_{21}(t_3)\sqrt{a_{21}^{\rm pr}}e^{i\phi_{21}^{\rm pr}}$ at $t=t_3$. Here, subscripts represent the quantum states ($mn$ for transition $|n\rangle\rightarrow |m\rangle$), and superscripts denote the pulse ("pr" for the probe pulse and "pu" for the pump pulse). Similarly, the TDM $\mu_{31}(t)$ becomes $\mu_{31}(0)\sqrt{a_{31}^{\rm pu}}e^{i\phi_{31}^{\rm pu}}$ at $t=0$, $\mu_{31}(t_1)\sqrt{a_{31}^{\rm pu}}e^{-i\phi_{31}^{\rm pu}}$ at $t=t_1$, and $\mu_{31}(t_3)\sqrt{a_{31}^{\rm pr}}e^{-i\phi_{31}^{\rm pr}}$ at $t=t_3$. In the pump--probe scenario, $t_1=0$, $t_2=\tau$, and $t_3=t$ will be used. During the interval between interactions, i.e., the field-free evolution, population relaxation occurs at a rate of $\Gamma_{21}$ ($\Gamma_{31}$) for diagonal element $\rho_{22}$ ($\rho_{33}$) of the density matrix, and dephasing occurs at a rate of $\gamma_{21}=\frac{\Gamma_{21}}{2}$ ($\gamma_{31}=\frac{\Gamma_{31}}{2}$) for nondiagonal element $\rho_{21}$ ($\rho_{31}$) of the density matrix; $\gamma_{32}=\frac{\Gamma_{21}+\Gamma_{31}}{2}$ is responsible for dephasing of nondiagonal elements $\rho_{32}$ and $\rho_{23}$. The response function of the stimulation process is
\begin{equation}
\begin{split}
R_{\rm st}(t,\tau,0)\propto & ~ia_{21}^{\rm pu}\sqrt{a_{21}^{\rm pr}}e^{i\phi_{21}^{\rm pr}}\mu_{21}^4e^{-\Gamma_{21}\tau}e^{-i\omega_{21}t-\gamma_{21}t}\\
&+ia_{31}^{\rm pu}\sqrt{a_{31}^{\rm pr}}e^{i\phi_{31}^{\rm pr}}\mu_{31}^4e^{-\Gamma_{31} \tau}e^{-i\omega_{31}t-\gamma_{31}t},
\end{split}
\end{equation}
in which the time-dependent TDMs $\mu_{21}(t)$ and $\mu_{31}(t)$ are approximated as constant~\cite{PeterHammSpec}. The phases $e^{i\phi_{21}^{\rm pu}}$ and $e^{-i\phi_{21}^{\rm pu}}$ in the modification factors cancel each other for $\mu_{21}(t)$, and it is the same for $\mu_{31}(t)$. 

Following the same procedure, we can obtain the response function of the bleach and quantum beat pathways in the pump--probe scenario. All these response functions are given by
\begin{equation}
\begin{split}
&R_{\rm st}(t,\tau,0)\propto\\
&i\,a_{21}e^{i\phi_{21}^{\rm pr}}\mu_{21}^4e^{-\Gamma_{21}\tau}e^{-i\omega_{21}t-\gamma_{21}t}+\\
&ia_{31}e^{i\phi_{31}^{\rm pr}}\mu_{31}^4e^{-\Gamma_{31}\tau}e^{-i\omega_{31}t-\gamma_{31}t},\\
&R_{\rm bl}(t,\tau,0)\propto\\
&i\,a_{21}e^{i\phi_{21}^{\rm pr}}\mu_{21}^4e^{-i\omega_{21}t-\gamma_{21}t}+ia_{31}e^{i\phi_{31}^{\rm pr}}\mu_{31}^4e^{-i\omega_{31}t-\gamma_{31}t},\\
&R_{\rm qb}(t,\tau,0)\propto\\
&i\,b(I_{\rm pr})\sqrt{a_{21}a_{31}}e^{i(\phi_{31}^{\rm pr}-\delta)}\mu_{21}^2\mu_{31}^2e^{i\omega_{32}\tau}e^{-\Gamma_{32} \tau}e^{-i\omega_{21}t-\gamma_{21}t}+\\
&i\frac{1}{b(I_{\rm pr})}\sqrt{a_{21}a_{31}}e^{i(\phi_{21}^{\rm pr}+\delta)}\mu_{21}^2\mu_{31}^2e^{-i\omega_{32}\tau}e^{-\Gamma_{32} \tau}e^{-i\omega_{31}t-\gamma_{31}t}.
\end{split}
\end{equation}
Here, $a_{21}=a_{21}^{\rm pu}\sqrt{a_{21}^{\rm pr}}$, $a_{31}=a_{31}^{\rm pu}\sqrt{a_{31}^{\rm pr}}$, $\delta=(\phi_{31}^{\rm pu}-\phi_{21}^{\rm pu})$, and $b(I_{\rm pr})=\sqrt{\frac{a_{31}^{\rm pr}}{a_{21}^{\rm pr}}}$. The subscripts "st", "bl" and "qb" represent the stimulation, bleach, and quantum beat pathways, respectively. 

The absorption spectrum is the Fourier transform of the total response function, i.e., $R_{\rm st}+R_{\rm bl}+R_{\rm qb}$, and it reads as
\begin{equation}
\begin{split}
S\left(\omega,\tau\right)&\propto {\rm Im}\{-i\frac{a_{21}e^{i\phi_{21}^{\rm pr}}(\frac{1}{2}e^{-\Gamma_{21}\tau}+\frac{1}{2})}{i(\omega-\omega_{21})-\gamma_{21}}\\
&-i\frac{b(I_{\rm pr})\sqrt{a_{21}a_{31}}e^{i(\phi_{31}^{\rm pr}-\delta)}e^{i\omega_{32}\tau}e^{-\Gamma_{32} \tau}}{i(\omega-\omega_{21})-\gamma_{21}}\\
&-i\frac{a_{31}e^{i\phi_{31}^{\rm pr}}(2e^{-\Gamma_{31} \tau}+2)}{i(\omega-\omega_{31})-\gamma_{31}}\\
&-i\frac{\frac{1}{b(I_{\rm pr})}\sqrt{a_{21}a_{31}}e^{i(\phi_{21}^{\rm pr}+\delta)}e^{-i\omega_{32}\tau}e^{-\Gamma_{32} \tau}}{i(\omega-\omega_{31})-\gamma_{31}}
\},
\end{split}
\label{equ:puprsf}
\end{equation}
where $\mu_{31}=\sqrt{2}\mu_{21}$~\cite{TDM} is used.

In the probe--pump scenario, there is a phase lag $e^{i\omega \tau}$ between the emitted field and probe pulse, because the probe pulse comes first, but emission of the third-order polarization starts only after the last interaction. This phase lag will cause time-dependent patterns in the absorption spectrum, which are referred to as perturbed free induction decay~\cite{PFID1,PFID2,PFID3,PFID4}. In the probe--pump scenario, $t_1=\tau$, $t_2=0$, and $t_3=t$ will be used. The absorption spectrum in the probe--pump scenario can be given similarly to that in the pump--probe scenario:
\begin{equation}
\begin{split}
S(\omega,\tau)&\propto {\rm Im}\{e^{i\omega\tau} [-i\frac{a_{21}e^{i\phi_{21}^{\rm pr}}e^{-i\omega_{21}\tau-\gamma_{21}\tau}}{i(\omega-\omega_{21})-\gamma_{21}}\\
&-i\frac{b(I_{\rm pr})\sqrt{a_{21}a_{31}}e^{i(\phi_{31}^{\rm pr}-\delta)}e^{-i\omega_{31}\tau-\gamma_{31}\tau}}{i(\omega-\omega_{21})-\gamma_{21}}\\
&-i\frac{4a_{31}e^{i\phi_{31}^{\rm pr}}e^{-i\omega_{31}\tau-\gamma_{31}\tau}}{i(\omega-\omega_{31})-\gamma_{31}}\\
&-i\frac{\frac{1}{b(I_{\rm pr})}\sqrt{a_{21}a_{31}}e^{i(\phi_{21}^{\rm pr}+\delta)}e^{-i\omega_{21}\tau-\gamma_{21}\tau}}{i(\omega-\omega_{31})-\gamma_{31}}]
\}.
\end{split}
\label{equ:prpusf}
\end{equation}
The variable substitutions are the same as those in \eqref{equ:puprsf}. 

For convenience, the value of $b(I_{\rm pr})$ is approximated as $b(I_{\rm pr})=\sqrt{\frac{\mu_{31}}{\mu_{21}}}$. The undetermined parameters, $a_{21}$, $a_{31}$, $\phi_{21}^{\rm pr}$, $\phi_{31}^{\rm pr}$, and $\delta$, can be extracted from the experimental absorption spectrum. The parameter $a_{21}$ ($a_{31}$) represents the total amplitude modification of $\mu_{21}$ ($\mu_{31}$) induced by both the pump and probe pulses. The parameter $\phi_{21}^{\rm pr}$ ($\phi_{31}^{\rm pr}$) represents the phase modification induced by the probe pulse, and $\delta$ represents the modification contributed by the pump pulse in the quantum beat pathway. In the process of extracting the modulation factors, $\delta=\phi_{31}^{\rm pu}-\phi_{21}^{\rm pu}$ is set to zero. The effect of $\delta$ is distributed in $\phi_{21}^{\rm pr}$ and $\phi_{31}^{\rm pr}$, as the modulation effect induced by the pump pulse denoted by the amplitude factors has been included in $a_{21}$ and $a_{31}$, and the phase and amplitude factors are related to each other.
\begin{figure}[ht]
\centering
\includegraphics[width=3.4in]{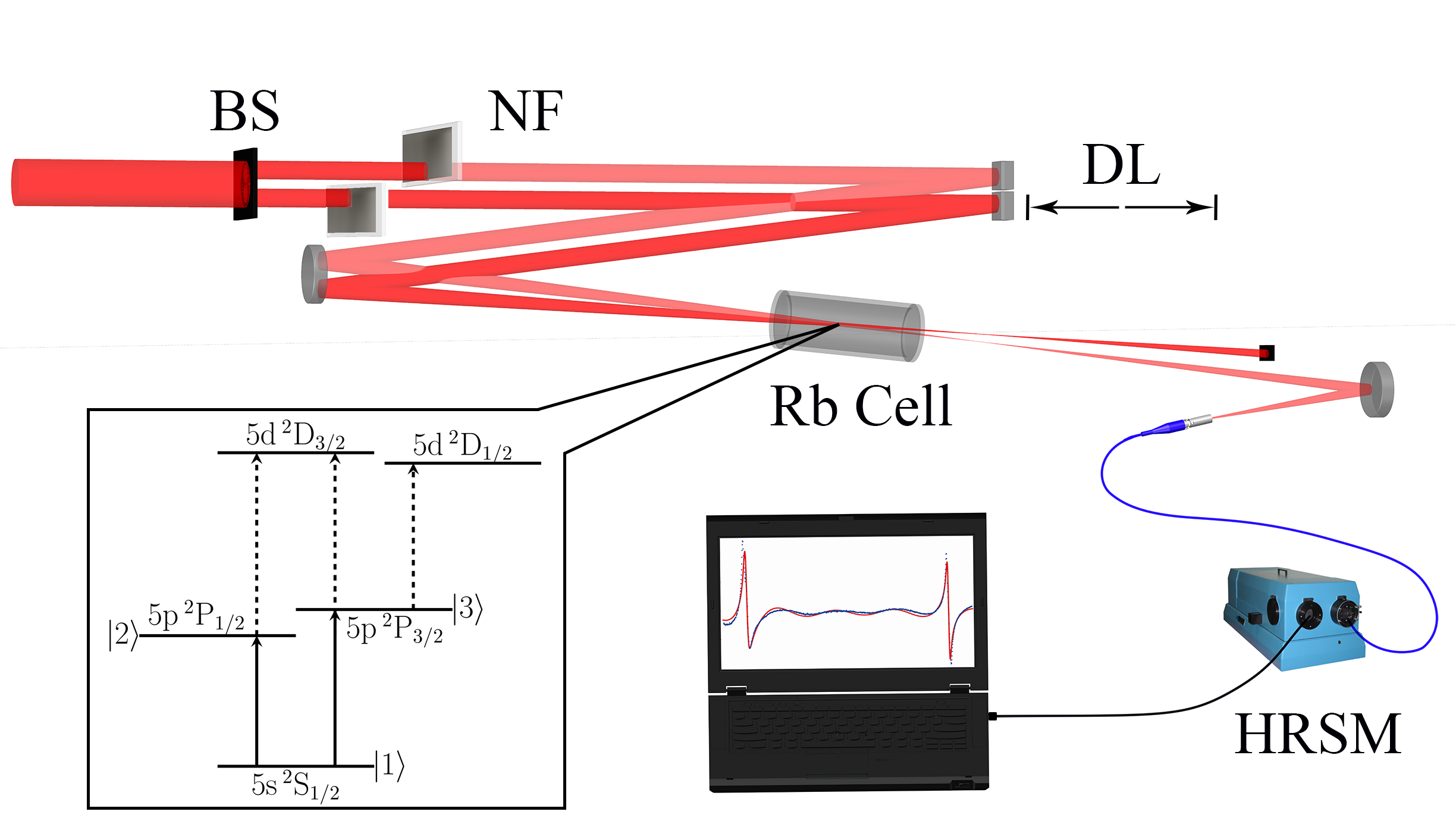}
\caption{Sketch of the experimental setup for the pump--probe measurement; inset shows the configuration of the three-level system in atomic rubidium. BS, beam splitter. NF, neutral density filter. DL, delay line. HRSM, high-resolution spectrometer.}
\label{fig:setup}
\end{figure}

\section{Experiment setup}

The pump--probe experimental setup is shown in Fig.\,\ref{fig:setup}. Except for the neutral density filter, which is used to adjust the laser field intensity, all the optical elements are reflective. Rubidium is used as the sample in the measurement and is contained in a 2.5-cm-long cell with 1.5-mm-thick windows of BK7 glass. Due to the craft, there is background nitrogen gas at a pressure of $5.0 \times 10^{-5}~{\rm Pa}$ in the cell at room temperature. During the measurement, the temperature of the Rb cell is held at $135.0 \pm 2.0~^\circ$C by a home-made oven, and the corresponding atom density is estimated to be $(1.8 \pm 0.2) \times 10^{14}~{\rm cm}^{-3}$~\cite{rubidium}.

Fourier-transform-limited laser fields (30\,fs, 1\,kHz, 800\,nm, FWHM of 20\,nm) are supplied by a commercial Ti:sapphire laser system. The pump and probe beams are obtained by dividing the wave front of the main beam using a spatial mask with two irises, which irradiate two side-by-side mirrors that are separately mounted on two translation stages. These two mirrors can be moved backward and forward to scan the time delay $\tau$ over several picoseconds. In the measurement, the time delay ranges from -960\,fs to 960\,fs in 4.8\,fs steps. The pulse energies of the pump and probe beams can be adjusted separately by the continuously variable neutral density filters, and both pulses are focused into the rubidium cell by a concave mirror. After the interaction, the transmitted probe pulse is collected by a high-resolution spectrometer (McPherson 2061), which is set to cover only the excited states $5{\rm p}\,^2{\rm P}_{1/2}$ and $5{\rm p}\,^2{\rm P}_{3/2}$ in rubidium with a resolution of 0.035\,meV to obtain a high resonant-to-nonresonant ratio. During the measurement, the integral time of the spectrometer is set to 100\,ms with an average of 10 times, i.e., the measured absorption spectrum is averaged over $10^3$ pulses.

\begin{figure}[htbp]
\includegraphics[width=3.4in]{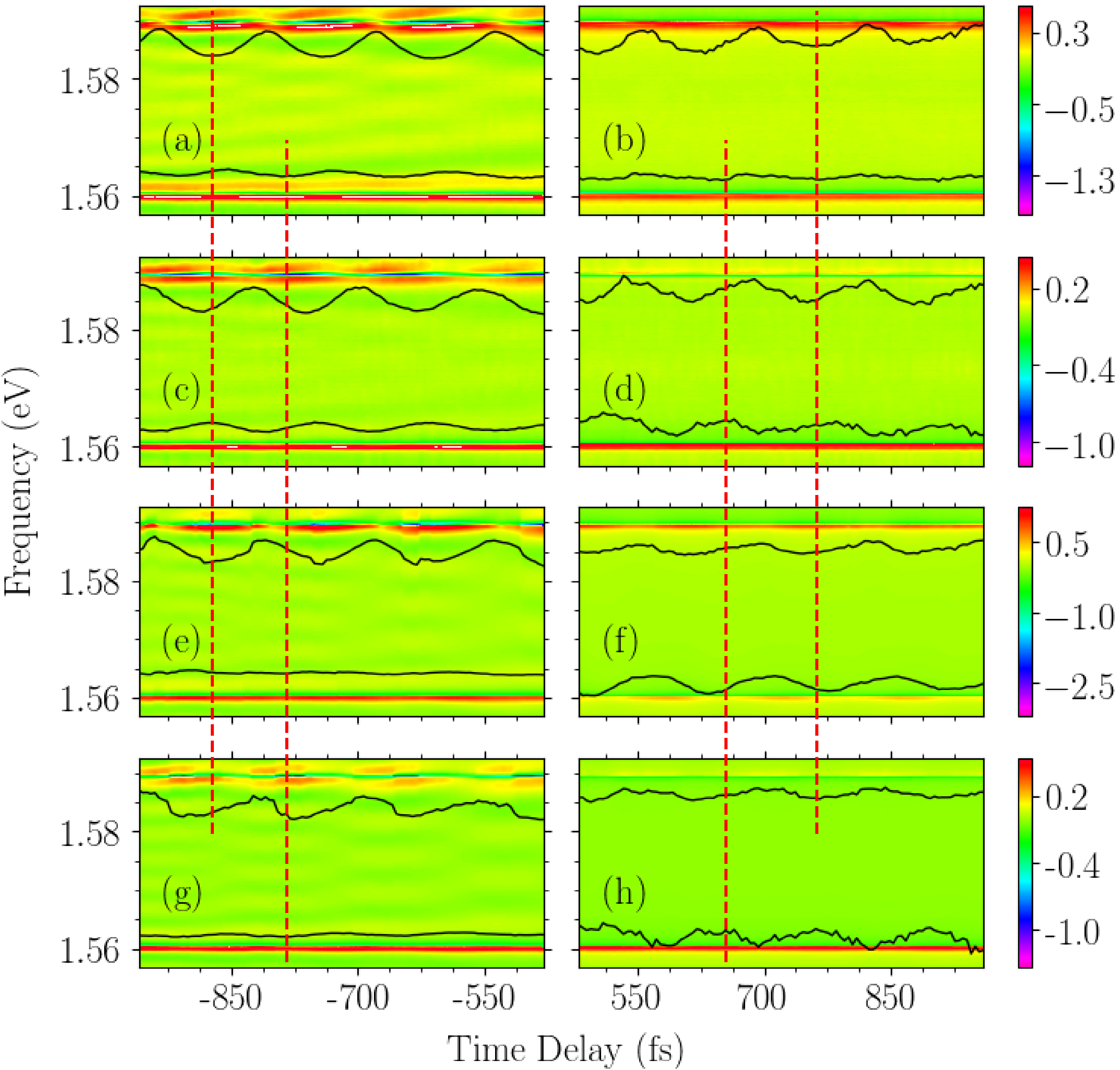}
\caption{Time-dependent transient absorption spectra collected with a fixed pump-pulse intensity of $I_{\rm pu}=2.0\times 10^{10}\,{\rm W/cm^2}$ and a probe-pulse intensity $I_{\rm pr}=0.8\times 10^{10}\,{\rm W/cm^2}$ for (a), (b) and $I_{\rm pr}=3.2\times 10^{10}\,{\rm W/cm^2}$ for (c), (d). (e), (f), (g) and (h) are reconstructed spectra for (a), (b), (c) and (d), respectively. Results are shown only in time delay ranges of -960 to -480\,fs and 480 to 960\,fs, where the pump and probe pulses are entirely separate. Oscillations induced by quantum beat at both transition energies are indicated by black lines.}
\label{fig:compare}
\end{figure}

\section{Results and discussion}

To study the dynamic evolution of the multilevel system in medium-strength fields, we perform transient absorption measurement in gaseous rubidium. The pump-pulse intensity is held at $2.0\times 10^{10}\,{\rm W/cm^2}$ in the measurement. The results for probe-pulse intensities $I_{\rm pr}$ of $0.8\times 10^{10}$ and $3.2\times 10^{10}\,{\rm W/cm^2}$ are shown in Figs.~\ref{fig:compare}(a), \ref{fig:compare}(b) and \ref{fig:compare}(c), \ref{fig:compare}(d), respectively. PFID is obvious when pump pulse follows the probe pulse. The line shape of the absorption spectra for transitions $5{\rm s}\,^2{\rm S}_{1/2}\rightarrow 5{\rm p}\,^2{\rm P}_{1/2,\,3/2}$ is deformed and is shown as the superposition of the emission and absorption at different energies owing to the modulation effect of the strong laser fields. The deformed line shape varies periodically with the pump--probe time delay $\tau$, and the changes are different in the probe--pump and pump--probe scenarios. Two time-dependent absorption lines centered on the transition energies $\omega_{21}$ = 1.56\,eV and $\omega_{31}$ = 1.59\,eV are selected and shown as black lines to emphasize the spectral evolution. Their shape and amplitude are both modulated as a function of the pump--probe time delay $\tau$ and feature oscillations with a period of $2\pi/\omega_{32} \approx 140\,{\rm fs}$. The shape of the time-dependent oscillations changes and the overall amplitude decreases when the probe-pulse intensity $I_{\mathrm{pr}}$ is tuned from $0.8\times 10^{10}\,{\rm W/cm^2}$ to $3.2\times 10^{10}\,{\rm W/cm^2}$. The time-dependent oscillations at both transition energies exhibit a phase shift regardless of which pulse interacts with the sample first. However, the magnitudes of the phase shifts are different, as demonstrated by the vertical lines that highlight the positions of the oscillations' minima. In the probe--pump scenario, the time-dependent oscillation at $\omega_{21}$ has a phase change of 0.34$\pi$ ($24\,{\rm fs}/140\,{\rm fs}\times 2\pi$), whereas that of the oscillation at $\omega_{31}$ is 0.36$\pi$, i.e., the phases of the oscillations at $\omega_{21}$ and $\omega_{31}$ change by approximately the same amount. In contrast, the phase change of the time-dependent oscillation at $\omega_{21}$ is approximately 0.82$\pi$, and that of the oscillation at $\omega_{31}$ is approximately 0.28$\pi$ in the pump--probe scenario, i.e. the phase at $\omega_{21}$ changes by roughly $\pi/2$ more than that at $\omega_{31}$.

\begin{figure}[htbp]
\centering
\includegraphics[width=3.4in]{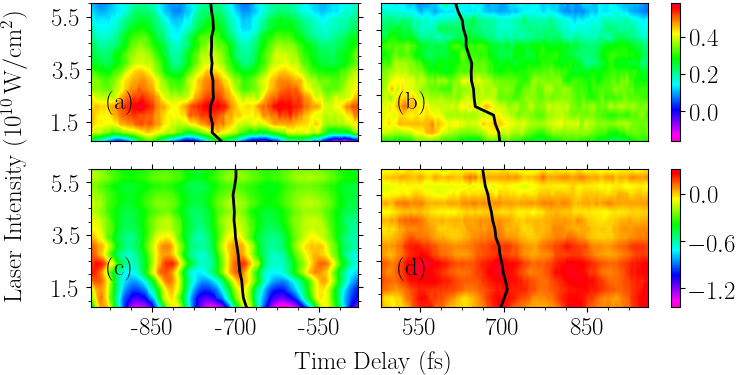}
\caption{Time-dependent absorption spectra at transition energies $\omega_{21}$ = 1.56\,eV and $\omega_{31}$ = 1.59\,eV as a function of the probe-pulse intensity for transition $|1\rangle \rightarrow |2\rangle$ in the (a) probe--pump and (b) pump--probe scenarios and for transition $|1\rangle \rightarrow |3\rangle$ in the (c) probe--pump and (d) pump--probe scenarios. The black solid line in each panel corresponds to the maximum of the absorption spectrum.}
\label{fig:qbphase}
\end{figure}

\begin{figure*}[htbp]
\centering
\includegraphics[width=7in]{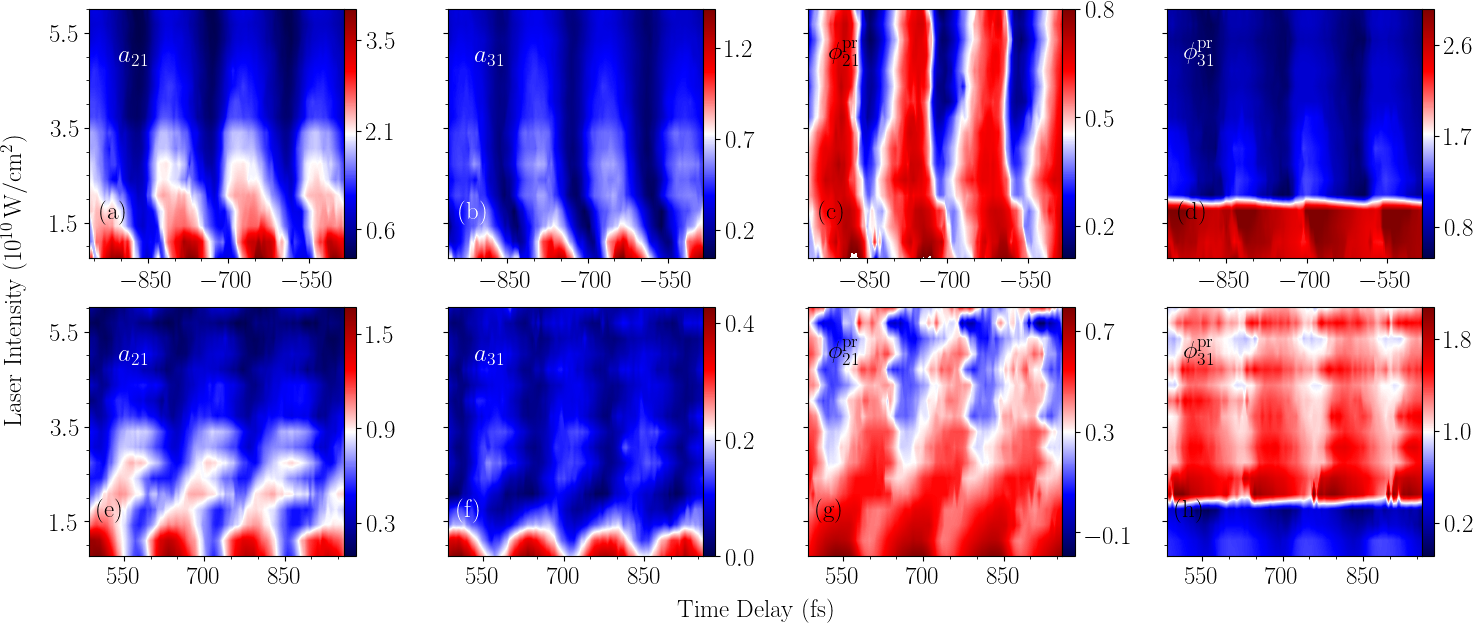}
\caption{Time-dependent amplitude and phase modulation factors as functions of the probe-pulse intensity with the pump-pulse intensity fixed at $I_{\rm pu}=2.0\times 10^{10}\,{\rm W/cm^2}$. (a) $a_{21}$, (b) $a_{31}$, (c) $\phi_{21}^{\rm pr}$, and (d) $\phi_{31}^{\rm pr}$ for the pump--probe scenario. (e), (f), (g), and (h) Results for the probe--pump scenario. Here $a_{21}$ and $a_{31}$ are given in arbitrary units, and the unit of $\phi_{21}^{\rm pr}$ and $\phi_{31}^{\rm pr}$ is $\pi$. The black line in each panel show the maximum of the corresponding modulation factor as a function of the probe-pulse intensity.}
\label{fig:fitpara}
\end{figure*}

The different results in the probe--pump and pump--probe scenarios demonstrate the different dynamic evolution of the sample with the different arrival order of the pump and probe pulses. As the phase of the time-dependent oscillation can reflect the phase of the sample's polarization, it was used to reconstruct the atomic system's phase information~\cite{spspLiu1}. Because medium-strength laser pulses are applied in these measurements, it becomes more difficult to recognize and understand the observed spectral features, and the well-known perturbation theory is no longer suitable. Rabi oscillation can no longer be ignored. Further, the strong laser fields can also include higher excited states in the interaction, e.g., $5\mathrm{d}\,^{2}\mathrm{D}_{3/2,\,5/2}$, as illustrated in Fig.~\ref{fig:setup}, which may affect the dynamic evolution's amplitude of the observed V-type three-level system in the measurement but have no influence on the phase. However, the intensity of the laser pulse is still not high enough to cause strong field ionization. The effect of strong laser fields on the multilevel system results in nontrivial changes in the populations of quantum states and the coherence between quantum states, which go beyond those described by perturbation theory. The corresponding dynamics is reflected in the time-dependent transient absorption spectra. The dynamic evolution of the multilevel system in such laser fields is described by~\cref{equ:puprsf,equ:prpusf}, which are developed from perturbation theory with a modified TDM. The time-dependent modification factors can be obtained by fitting the experimental data to \eqref{equ:puprsf} in the pump--probe scenario and \eqref{equ:prpusf} in the probe--pump scenario. The residual sum of squares according to the least square method, 
\begin{equation}
\chi^{2}=\sum_i {[S(\omega_i,\tau)-S_{\mathrm{exp}}(\omega_i,\tau)]^2},
\label{eq:7th}
\end{equation}
is applied to evaluate these modulation factors, $a_{21}$, $a_{31}$, $\phi_{21}^{\rm pr}$, and $\phi_{31}^{\rm pr}$, that are extracted from the experimental spectrum at a fixed time delay $\tau$. Here, $S(\omega_i,\,\tau)$ is the simulated absorption spectrum calculated at $\omega_i$ by~\eqref{equ:puprsf} in the pump--probe scenario or \eqref{equ:prpusf} in the probe--pump scenario, and $S_{\mathrm{exp}}(\omega,\tau)$ is the corresponding experimental result. The sum is taken over all the elemental frequencies in the observed spectra. A set of amplitude and phase modulation factors is obtained that corresponds to the minimum residual sum of squares $\chi^{2}$ at the fixed time delay $\tau$. By introducing these modification factors into ~\cref{equ:puprsf,equ:prpusf}, the time-dependent absorption spectra corresponding to probe-pulse intensities $I_{\rm pr}=0.8\times 10^{10}\,{\rm W/cm^2}$ and $I_{\rm pr}=3.2\times 10^{10}\,{\rm W/cm^2}$ are calculated, as shown in Figs.~\ref{fig:compare}(c), \ref{fig:compare}(d) and \ref{fig:compare}(e), \ref{fig:compare}(f), respectively. Oscillations at both transition energies are also indicated by black lines, and they are in phase with the experimental results in both the pump--probe and probe--pump scenarios, as illustrated by the vertical lines in Fig.~\ref{fig:compare}, demonstrating the validity of the theoretical model. These amplitude and phase modification factors also provide a method to quantify the modulation effect due to intense laser fields. 

Transient absorption measurement in which the probe-pulse intensity $I_{\rm pr}$ is continuously changed from $0.8\times 10^{10}\,{\rm W/cm^2}$ to $6.0\times 10^{10}\,{\rm W/cm^2}$ is performed to further examine the dynamic evolution of the multilevel system in medium-strength laser fields. The measured absorption spectra reveal that both the absorption line shape at fixed pump--probe time delays and the time-dependent spectra at transition energies change with the probe-pulse intensity. The absorption spectra at transition energies of $\omega_{21}$ and $\omega_{31}$ measured at varying probe-pulse intensities are displayed in Figs.~\ref{fig:qbphase}(a), \ref{fig:qbphase}(b) and \ref{fig:qbphase}(c), \ref{fig:qbphase}(d), respectively. The black lines associated with the maxima of the absorption spectra in each panel display the phase shift as a function of the probe-pulse intensity. In the probe--pump scenario, the peak of the oscillation at $\omega_{21}$ is steeper than that at $\omega_{31}$, i.e., the contrast in Fig.~\ref{fig:qbphase}(c) is better than that in Fig.~\ref{fig:qbphase}(a). Both of them shift toward negative time delays with increasing probe-pulse intensity, as illustrated by the black lines. The time-dependent oscillation at $\omega_{21}$ in the pump--probe scenario also shifts toward negative time delays with increasing probe-pulse intensity. There is a fast phase shift of approximately $\pi/2$ (35 fs) when the probe-pulse intensity is tuned from $1.7\times 10^{10}\,{\rm W/cm^2}$ to $2.0\times 10^{10}\,{\rm W/cm^2}$. However, the oscillation at $\omega_{31}$ in the pump--probe scenario first shifts toward positive time delays when the probe pulse intensity is changed from $0.8\times 10^{10}\,{\rm W/cm^2}$ to $1.7\times 10^{10}\,{\rm W/cm^2}$ and then shifts linearly in the opposite direction as the probe-pulse intensity increases further. This result differs from that of a previous study~\cite{spspLiu1}. When weak probe and strong pump pulses are used, the phases for both transitions shift in the same direction in the pump--probe scenario and in opposite directions in the probe--pump scenario owing to the increase in the pump-pulse intensity. This difference is caused by the application of an intense probe pulse in these measurements.
The amplitudes of both time-dependent oscillations first increase with the probe-pulse intensity. Then, as the probe-pulse intensity increases further, the intense laser pulse will couple the excited system to even higher states (e.g., $5\mathrm{d}\,^{2}\mathrm{D}_{3/2,\,5/2}$) when the probe-pulse intensity exceeds $I_{\rm pr}=2.2\times10^{10}\,{\rm W/cm^2}$, resulting in a decrease in the amplitudes, as shown in Fig.~\ref{fig:qbphase}.

By globally fitting the measured time-dependent absorption spectra to \eqref{equ:puprsf} in the pump--probe scenario and \eqref{equ:prpusf} in the probe--pump scenario, the time-dependent and intensity-dependent modulation factors for probe-pulse intensities varying from $0.8\times 10^{10}\,{\rm W/cm^2}$ to $6.0\times 10^{10}\,{\rm W/cm^2}$ are obtained. The results in Fig.~\ref{fig:fitpara} show the time-dependent amplitude factors $a_{21}$, $a_{31}$ and phase factors $\phi_{21}^{\rm pr}$, $\phi_{31}^{\rm pr}$ as a function of the probe-pulse intensity $I_{\rm{pr}}$ at a fixed pump-pulse intensity of $I_{\rm{pu}}=2.0\times 10^{10}\,{\rm W/cm^2}$. These factors change periodically with the same repetition period of 140\,fs owing to interference between the transitions $|1\rangle\rightarrow|2\rangle$ and $|1\rangle\rightarrow|3\rangle$. With increasing probe-pulse intensity, the maxima of the amplitude modulation factors $a_{21}$ and $a_{31}$ shift toward negative time delays in the probe--pump scenario and toward positive time delays in the pump--probe scenario. The maxima of the amplitude factors decrease with increasing probe-pulse intensity in both scenarios. For the phase modulation factors, the maximum of $\phi_{21}^{\rm pr}$ shifts toward negative time delays in the probe--pump scenario until $I_{\rm pr}= 3.0 \times 10^{10}\,{\rm W/cm^2}$ and toward positive time delays in the pump--probe scenario until $I_{\rm pr}=2.0\times 10^{10}\,{\rm W/cm^2}$, and then the shift of $\phi_{21}^{\rm pr}$ in both scenarios changes to approximately zero as $I_{\rm pr}$ increases further. The maximum of $\phi_{21}^{\rm pr}$ decreases with increasing probe-pulse intensity in both the probe--pump and pump--probe scenarios. The maximum $\phi_{31}^{\rm pr}$ value decreases rapidly by 1.5 in the probe--pump scenario, and that in the pump--probe scenario changes in just the opposite way when the probe-pulse intensity is tuned from $1.7\times 10^{10}\,{\rm W/cm^2}$ to $2.0\times 10^{10}\,{\rm W/cm^2}$.

\begin{figure*}[ht!]
\centering
\includegraphics[width=6.0 in]{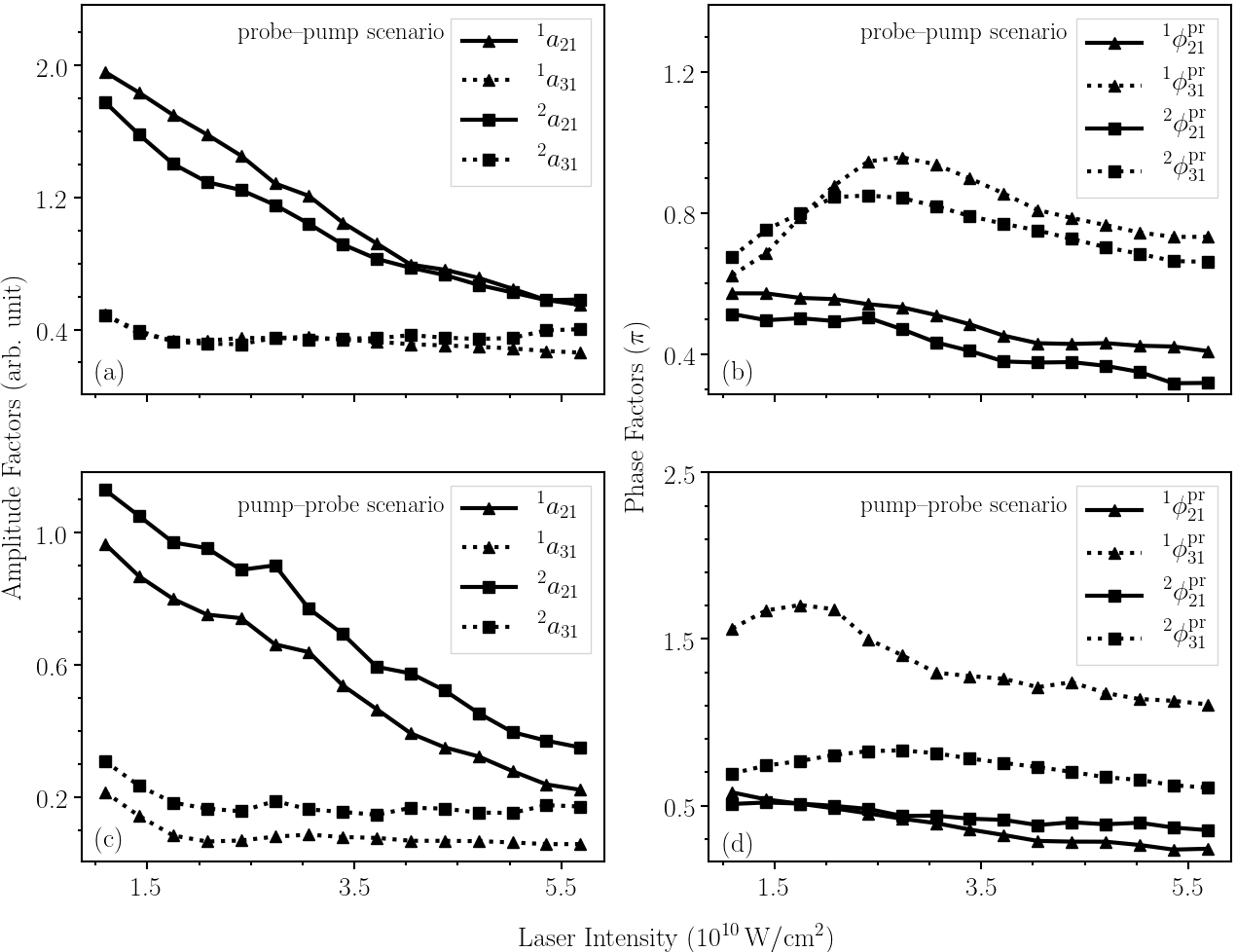}
\caption{Dependence of the averaged modification factors on the probe-pulse intensity at a pump-pulse intensity $I_{\rm{pu}}$ of $2.0\times 10^{10}\,{\rm W/cm^2}$ or $4.0\times 10^{10}\,{\rm W/cm^2}$. (a) Amplitude and (b) phase factors extracted from the measurement in the pump--probe scenario and (c) amplitude and (d) phase factors in the probe--pump scenario. The superscripts ``1'' and ``2'' on the left side of these modulation factors denote the results for $I_{\mathrm{pu}}=2.0\times 10^{10}\,{\rm W/cm^2}$ and $I_{\mathrm{pu}}=4.0\times 10^{10}\,{\rm W/cm^2}$, respectively.}
\label{fig:onecycle}
\end{figure*}

We have discussed the time-dependent properties of these modulation factors so far. To better explain this point and describe the dynamic modulation resulting from the presence of medium-strength laser fields, these four modulation factors are averaged over one cycle\,(140\,fs), in the probe--pump scenario or the pump--probe scenario to remove the influence of quantum beats. The results [Fig.~\ref{fig:onecycle}] illustrate the dependence of the averaged amplitude and phase modification factors on the probe-pulse intensity. These averaged modulation factors can be taken as the DC component of the time-dependent modification factors, which can more obviously represent the modulation to the TDM induced by strong laser fields. Instead of the fixed pump-pulse intensity $I_{\rm pu}=2.0\times 10^{10}\,{\rm W/cm^2}$ that we discussed previously, another pump-pulse intensity, $I_{\rm pu}=4.0\times 10^{10}\,{\rm W/cm^2}$, is applied to examine the effect of the pump pulse on the system's dynamics. To identify the modification factors of different pump-pulse intensities in each scenario, a superscript on the left side is introduced, where 1 and 2 denote the pump-pulse intensities $I_{\mathrm{pu}}=2.0\times 10^{10}\,{\rm W/cm^2}$ and $I_{\rm pu}=4.0\times 10^{10}\,{\rm W/cm^2}$, respectively. These averaged factors in the pump--probe and probe--pump scenarios change differently with increasing probe-pulse intensity.

The averaged amplitude factors for various probe-pulse intensities in the probe--pump and pump--probe scenarios are illustrated in Figs.~\ref{fig:onecycle}(a) and \ref{fig:onecycle}(c), respectively. The results show that $^1a_{21}$ and $^2a_{21}$ decrease almost linearly with the probe-pulse intensity regardless of whether the probe pulse precedes or follows the pump pulse in interacting with the sample, whereas $^1a_{31}$ and $^2a_{31}$ first decrease until $I_{\rm pr}=2.0\times 10^{10}\,\rm{W/cm^{2}}$ and then remain constant as the probe-pulse intensity increases further. Even though transitions to higher states of the rubidium atom, e.g., $5{\mathrm d}\,^2{\mathrm D}_{3/2}$ and $5{\mathrm d}\,^2{\mathrm D}_{5/2}$, are ignored in this theoretical model, their contributions are reflected by the amplitude modification factors since they have no influence on quantum beat's phase and phase modification factors. With increasing probe-pulse intensity, more populations will be coupled to the higher excited states owing to single photon absorption, resulting in the changes in the amplitude factors. The populations of states $5{\rm p}\,^2{\rm P}_{1/2}$ and $5{\rm p}\,^2{\rm P}_{3/2}$ can be translated to $5{\mathrm d}\,^2{\mathrm D}_{3/2}$ and $5{\mathrm d}\,^2{\mathrm D}_{3/2,\,5/2}$, respectively, through resonance photon absorption, as shown in Fig.\,\ref{fig:setup}. The TDMs of the transitions $5{\rm p}\,^2{\rm P}_{1/2} \rightarrow 5{\rm d}\,^2{\rm D}_{3/2}$ (1.63\,eV) are approximately half of the total TDMs of $5{\mathrm d}\,^2{\mathrm P}_{3/2} \rightarrow 5{\mathrm d}\,^2{\mathrm D}_{3/2,\,5/2}$ (1.60\,eV) \cite{RbTDM}, i.e., the transition strengths have a ratio of approximately 1:4 if the spectral intensities of the laser fields at the transition energies are identical. However, the spectrum of the laser pulse used in the measurement has a Gaussian distribution with a center frequency of 1.55\,eV, and the spectral intensity at 1.60\,eV is approximately twice that at 1.63\,eV. At a lower probe-pulse intensity, the spectral intensity at 1.60\,eV is high enough to couple the state $5{\mathrm p}\,^2{\mathrm P}_{3/2}$ to $5{\mathrm d}\,^2{\mathrm D}_{3/2,\,5/2}$; hence, smaller amplitude factors $^1a_{31}$ and $^2a_{31}$ are extracted from the measurement. In contrast, the amplitude factors $^1a_{21}$ and $^2a_{21}$ are larger, as the spectral intensity at 1.63\,eV is fairly low. The resonance coupling effect for the transition $5{\rm p}\,^2{\rm P}_{1/2} \rightarrow 5{\rm d}\,^2{\rm D}_{3/2}$ becomes stronger with increasing probe-pulse intensity, causing the obvious decrease in the modulation factors $^1a_{21}$ and $^2a_{21}$. From the evolution of the averaged factors $^1a_{31}$ and $^2a_{31}$, one can find that the coupling effect reaches saturation with increasing probe-pulse intensity, i.e., saturation of the modulation effect to the state $5{\rm p}\,^2{\rm P}_{3/2}$. It was reported previously on the basis of near-infrared (NIR)--extreme ultraviolet measurement that an intense NIR laser pulse can stop the exponential dipole decay of a doubly excited state through complete ionization of the excited state~\cite{Science.354.738}, i.e., the amplitude of the electric dipole is held at zero by the intense NIR pulse. However, the results are different here, and the amplitude factors are always larger than zero. Resonance coupling to higher excited states cannot prevent coherence between quantum states $5{\rm p}\,^2{\rm P}_{1/2}$ and $5{\rm p}\,^2{\rm P}_{3/2}$ caused by quantum beats between these two excited states. The results obtained for two different pump-pulse intensities are compared to check the influence of the pump-pulse intensity on the response of the sample. When the probe-pulse intensity is less than $3.7\times 10^{10}\,{\rm W/cm^2}$, $^1a_{21}$ is larger than $^2a_{21}$, and $^1a_{31}$ is almost equal to $^2a_{\,31}$ in the probe--pump scenario. As the probe-pulse intensity increases further, $^1a_{31}$ becomes smaller than $^2a_{31}$, and $^1a_{21}$ and $^2a_{21}$ have similar values. The results show that the pump-pulse intensity affects the response of state $5{\rm p}\,^2{\rm P}_{3/2}$ at low probe intensities and the response of state $5{\rm p}\,^2{\rm P}_{1/2}$ at higher probe intensities in the probe--pump scenario. However, the results in the pump--probe scenario are relatively simple. The difference $(^2a_{21}-^1a_{21})$ remains constant ($0.2$) at various probe-pulse intensities, and $^2 a_{31}$ is smaller than $^1 a_{31}$ by a fixed difference of $0.1$. That is, the influence of the pump pulse on the dynamic evolution of the multilevel system does not change with variation in the probe-pulse intensity in the pump--probe scenario.

The dependence of the averaged modulation phase factors on the probe-pulse intensity in both scenarios is also shown in Figs.~\ref{fig:onecycle}(b) and \ref{fig:onecycle}(d). These results differ from those obtained in a measurement with weak probe and strong pump pulses~\cite{fanoprofile,Proc.Natl.Acad.Sci.USA.112.15613}, which showed that the phase change of an atomic or molecular system is proportional to the laser-pulse intensity. Fig.~\ref{fig:onecycle}(b) displays the results in the probe--pump scenario. $^1\phi_{21}$ and $^2\phi_{21}$ change similarly with increasing probe-pulse intensity, and the difference between them remains around 0.2. For $^1\phi_{31}^{\rm pr}$ and $^2\phi_{31}^{\rm pr}$, they first increase to their maxima at $I_{\rm pr}=2.4\times 10^{10}\,{\rm W/cm^2}$ and $I_{\rm pr}=2.7\times 10^{10}\,{\rm W/cm^2}$, respectively, and then decrease as $I_{\rm pr}$ increases further. The difference $(^1\phi_{31}^{\rm pr}- ^2\phi_{31}^{\rm pr})$ is fixed at approximately 0.2 if the probe-pulse intensity is larger than $3.7\times 10^{10}\,{\rm W/cm^2}$. In contrast, the results in the pump--probe scenario are different. $^1\phi_{21}^{\rm pr}$ and $^1\phi_{21}^{\rm pr}$ have approximately the same value when the probe-pulse intensity is less than $2.7\times 10^{10}\,{\rm W/cm^2}$, and then the difference $(^2\phi_{21}^{\rm pr}-^1\phi_{21}^{\rm pr})$ increases from zero to 0.2. Further, $^1\phi_{31}^{\rm pr}$ and $^2\phi_{31}^{\rm pr}$ first increase to their maxima at $I_{\rm pr}=1.8\times 10^{10}\,{\rm W/cm^2} $ and $I_{\rm pr}=2.7 \times 10^{10}\,{\rm W/cm^2}$, respectively, and then decrease with increasing probe-pulse intensity. The difference $(^1\phi_{31}^{\rm pr}-^2\phi_{31}^{\rm pr})$ remains approximately $\pi/2$ when the probe-pulse intensity is tuned to be larger than $2.7\times 10^{10}\,{\rm W/cm^2}$. Comparing the averaged amplitude and phase factors measured under the same conditions, one can describe the response of the atomic system to strong laser fields. A larger amplitude factor corresponds to a smaller phase factor for the response of one excited state and vice versa; the amplitude and phase factors for one excited state are balanced regardless of which pulse first interacts with the sample.

\section{Conclusion}
In summary, we study the dynamic evolution of an atomic system in medium-strength laser fields by performing transient absorption measurement. Perturbation theory is generalized to the strong-pulse regime by using a modified TDM, which is applied to interpret the dynamic evolution of the V-type three-level system in rubidium. The undetermined parameters in the analytical model, i.e., the modification factors of the TDM, are extracted by fitting the experimental spectra to the model. The different behavior of the V-type three-level system in the pump--probe and probe--pump scenarios are the results of different interaction mechanisms. In the probe--pump scenario, the initial values of the phase and amplitude of the dipoles induced by the probe pulse are determined by the pulse area of the probe pulse, and both will be modified by the pump pulse. The pump pulse creates an initial state in the pump--probe scenario, which will affect the amplitude and phase of the dipole induced by the probe pulse. The general dynamic evolution of quantum systems in the strong-pulse regime can be understood easily in terms of these averaged amplitude and phase modulation factors. To realize full control, the central frequency and chirp of the laser pulse can be introduced as experimental variables to examine the corresponding dynamic evolution. The analytical model used here can also be easily extended to more complex systems to understand their ultrafast dynamics in the transition area between strong and weak fields.

\section*{Funding Information}
National Natural Science Foundation of China (11504148, 11135002); Fundamental Research Funds for the Central Universities (lzujbky-2016-35).
\bibliography{pra}
\end{document}